**ARTICLE**                                                                      **Open Access**

# Miniscope3D: optimized single-shot miniature 3D fluorescence microscopy

Kyrollos Yanny[1], Nick Antipa[2], William Liberti[2], Sam Dehaeck[3], Kristina Monakhova[2], Fanglin Linda Liu[2], Konlin Shen[2], Ren Ng[2] and Laura Waller[1,2]

## Abstract

Miniature fluorescence microscopes are a standard tool in systems biology. However, widefield miniature microscopes capture only 2D information, and modifications that enable 3D capabilities increase the size and weight and have poor resolution outside a narrow depth range. Here, we achieve the 3D capability by replacing the tube lens of a conventional 2D Miniscope with an optimized multifocal phase mask at the objective's aperture stop. Placing the phase mask at the aperture stop significantly reduces the size of the device, and varying the focal lengths enables a uniform resolution across a wide depth range. The phase mask encodes the 3D fluorescence intensity into a single 2D measurement, and the 3D volume is recovered by solving a sparsity-constrained inverse problem. We provide methods for designing and fabricating the phase mask and an efficient forward model that accounts for the field-varying aberrations in miniature objectives. We demonstrate a prototype that is 17 mm tall and weighs 2.5 grams, achieving 2.76 µm lateral, and 15 µm axial resolution across most of the $900 \times 700 \times 390 \ \mu m^3$ volume at 40 volumes per second. The performance is validated experimentally on resolution targets, dynamic biological samples, and mouse brain tissue. Compared with existing miniature single-shot volume-capture implementations, our system is smaller and lighter and achieves a more than 2× better lateral and axial resolution throughout a 10× larger usable depth range. Our microscope design provides single-shot 3D imaging for applications where a compact platform matters, such as volumetric neural imaging in freely moving animals and 3D motion studies of dynamic samples in incubators and lab-on-a-chip devices.

## Introduction

Miniature widefield fluorescence microscopes enable important applications in systems biology, for example, the optical recording of neural activity in freely moving animals[1–4] and long-term in situ imaging within incubators and lab-on-a-chip devices. These miniature microscopes, commonly called "Miniscopes," are developed by a vibrant open-source community[5] and made of 3D-printed parts and off-the-shelf components. Although the Miniscope is

Correspondence: Kyrollos Yanny (kyrollosyanny@gmail.com) or
Nick Antipa (naantipa@gmail.com)
[1]UCB/UCSF Joint Graduate Program in Bioengineering, University of California, Berkeley, CA 94720, USA
[2]Department of Electrical Engineering & Computer Sciences, University of California, Berkeley, CA 94720, USA
Full list of author information is available at the end of the article
These authors contributed equally: Kyrollos Yanny, Nick Antipa

designed for 2D fluorescence imaging only, many applications can benefit from imaging 3D structures.

Volumetric microscopy methods aim to capture 3D structures; however, they often rely on scanning (e.g., two-photon, light sheet), which is difficult to miniaturize and must strike a balance between temporal resolution and field-of-view (FoV). Two-photon microscopes have been implemented in small form factors[6,7], giving high resolution at the cost of motion artefacts[8], a limited FoV, and expensive hardware. Miniaturized light sheet microscopes achieve faster capture[9] but also depend on scanning, which causes motion artefacts and increases the complexity and size of the hardware.

Unlike scanning approaches, single-shot methods[10–17] offer faster capture speeds, with a temporal resolution limited only by the camera frame rate. These methods encode information from the entire volume into a 2D measurement and then computationally reconstruct the 3D information. Single-shot 3D fluorescence capture has been demonstrated using a lensless architecture[13,14] but lacks the integrated illumination that is required for in vivo imaging. In addition, these mask-only systems have no magnifying optics and thus are limited to a low effective numerical aperture (NA), resulting in poor lateral and axial resolutions. Other recent work combines coding elements with multi-fiber endoscopes to achieve single-shot non-fluorescence 3D, with relatively low resolution[18]. Recently, the miniature light-field microscope (MiniLFM)[19] demonstrated an integrated 3D fluorescence system with computationally efficient temporal video processing for neural activity tracking[20]. This system adds a standard microlens array (regularly spaced, unifocal) to the image plane of the Miniscope, giving it single-shot 3D capabilities at the cost of degraded lateral resolution and a larger and heavier device. The MiniLFM algorithm[20] is optimized for neural activity tracking applications and uses temporal video processing, which requires sparsity, multiple frames of capture, and a static structure in the sample.

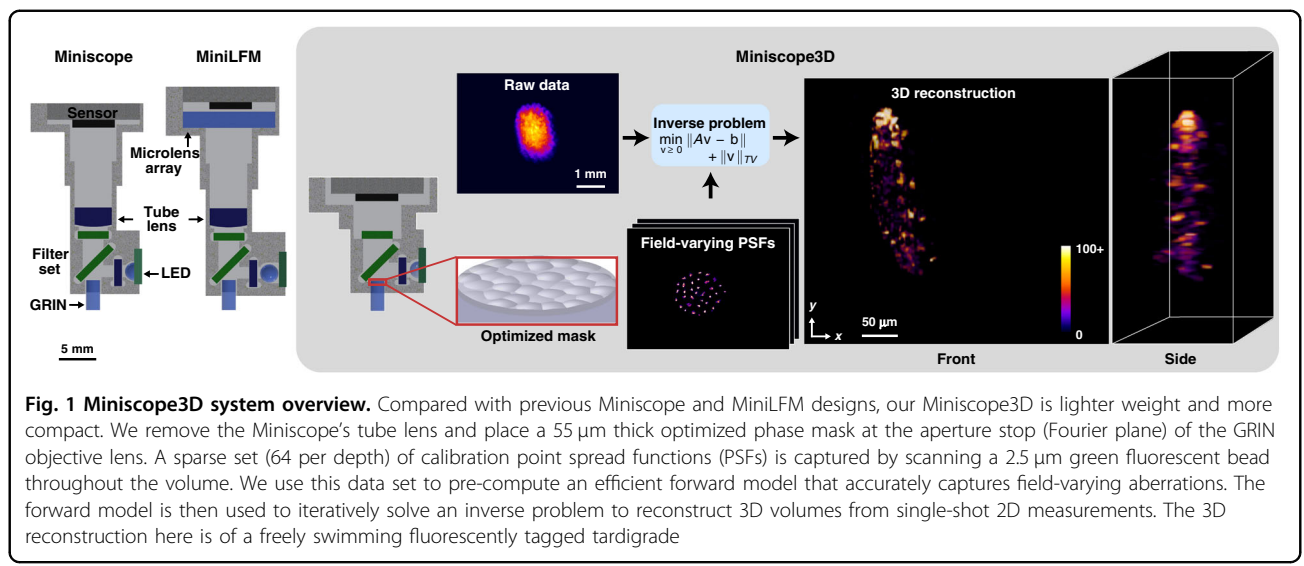

**Fig. 1 Miniscope3D system overview.** Compared with previous Miniscope and MiniLFM designs, our Miniscope3D is lighter weight and more compact. We remove the Miniscope's tube lens and place a 55 μm thick optimized phase mask at the aperture stop (Fourier plane) of the GRIN objective lens. A sparse set (64 per depth) of calibration point spread functions (PSFs) is captured by scanning a 2.5 μm green fluorescent bead throughout the volume. We use this data set to pre-compute an efficient forward model that accurately captures field-varying aberrations. The forward model is then used to iteratively solve an inverse problem to reconstruct 3D volumes from single-shot 2D measurements. The 3D reconstruction here is of a freely swimming fluorescently tagged tardigrade.

Here, we present a new single-shot 3D miniature fluorescence microscope, termed *Miniscope3D*, which is not only smaller and lighter weight than the MiniLFM but also achieves better resolution over a larger volume. It is designed as a simple hardware modification to the widely used UCLA Miniscope[5], replacing the tube lens with an optimized phase mask placed directly at the aperture stop (Fourier plane) of the objective lens (Fig. 1). The phase mask consists of a set of multifocal nonuniformly spaced microlenses, optimized such that each point within a 3D sample generates a unique high-frequency pattern on the sensor, encoding volumetric information in a single 2D measurement. The 3D volume is recovered by solving a sparsity-constrained compressed sensing inverse problem, enabling us to recover 24.5 million voxels from a 0.3 megapixel measurement. Our algorithm assumes the sample to be sparse in some domain, which is valid for a general class of fluorescent samples. We demonstrate the capabilities of our microscope by imaging fluorescent resolution targets, freely swimming biological samples, scattering mouse brain tissue, and optically cleared mouse brain tissue. We also validate the accuracy of our reconstructions against two-photon microscopy and discuss the limitations of our method.

To achieve high-quality imaging in a small, low-weight device, a number of technical innovations are developed. Placing the phase mask in Fourier space (instead of image space) significantly improves the compactness and reduces the computational burden[21-23]. Varying the focal lengths of the microlenses enhances the uniformity of the resolution across depths compared with implementations such as the MiniLFM. Because we use an optimized forward model and calibration scheme to account for the field-varying aberrations inherent to miniature objectives, we are able to add 3D capabilities to the 2D Miniscope at the cost of only a small loss in lateral resolution and a lower signal-to-noise ratio (SNR). Our algorithm unites optical theory with compressed sensing in a general way that can allow others to design and fabricate optimized phase masks for their applications. The main contributions of this work are as follows:

- A new miniature 3D microscope architecture that improves upon the MiniLFM, achieving significantly better resolution across a 10 × 10 larger usable deep while reducing the overall device size.
- A prototype, based on easily available parts, 3D printing, and open-source designs, that weighs

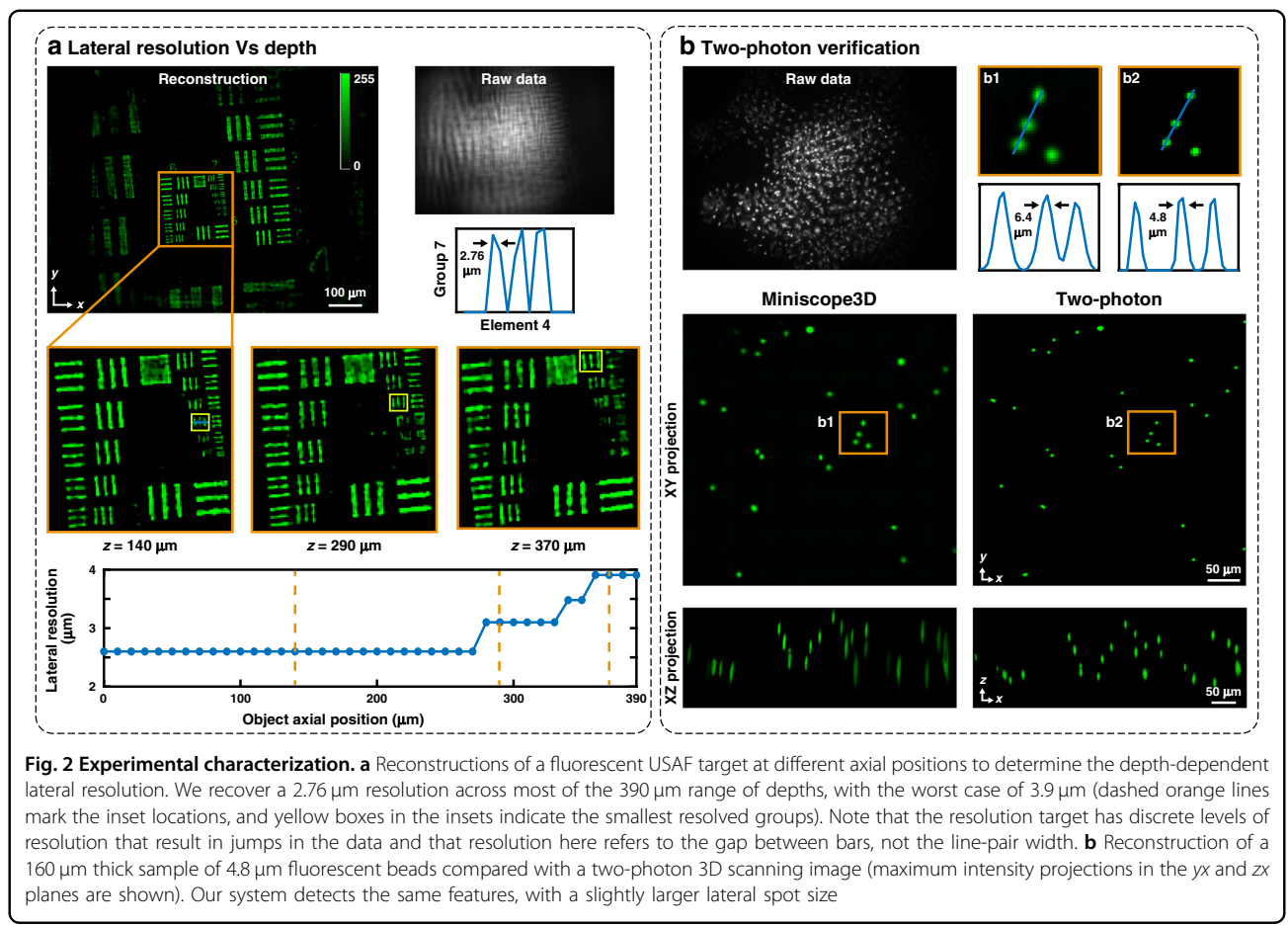

**Fig. 2 Experimental characterization. a** Reconstructions of a fluorescent USAF target at different axial positions to determine the depth-dependent lateral resolution. We recover a 2.76 μm resolution across most of the 390 μm range of depths, with the worst case of 3.9 μm (dashed orange lines mark the inset locations, and yellow boxes in the insets indicate the smallest resolved groups). Note that the resolution target has discrete levels of resolution that result in jumps in the data and that resolution here refers to the gap between bars, not the line-pair width. **b** Reconstruction of a 160 μm thick sample of 4.8 μm fluorescent beads compared with a two-photon 3D scanning image (maximum intensity projections in the *yx* and *zx* planes are shown). Our system detects the same features, with a slightly larger lateral spot size

2.5 grams and achieves 2.76 μm lateral and 15 μm axial resolution across most of the 900 × 700 × 390 μm³ volume at 40 volumes per second.

- Design principles for optimizing phase masks for 3D imaging and a high-quality fabrication method using two-photon polymerization in a Nanoscribe 3D printer.
- An efficient calibration scheme and reconstruction algorithm that accounts for the field-varying aberrations inherent in miniaturized objective lenses.

## Results

We characterize the performance of our computational microscope with samples of increasing complexity, capturing dynamic 3D recordings at frame rates of up to 40 volumes per second.

### Resolution characterization

The lateral resolution is measured at different depths by imaging a fluorescent resolution target every 10 μm axially and determining the smallest resolved group by eye. Figure 2a demonstrates a 2.76 μm uniform lateral resolution over 270 μm in depth. The resolution degrades to 3.9 μm over the next 120 μm in depth, for a total usable depth range of 390 μm. This relatively uniform resolution over a wide depth range is due to our multifocal design. Axial resolution is determined by imaging a thin layer of 4.8 μm fluorescent beads at different depths and using the Rayleigh criterion (at least a 20% dip between the peaks of the two reconstructed points) to determine the resolution. Raw data from multiple depths are added to synthesize a measurement of two layers of beads with varying separations (see Supplementary Fig. 2). We achieve a 15 μm axial resolution across the entire 390 μm depth range, which matches the axial full-width-half-maximum in the reconstructions of the 3D fluorescent bead sample in Fig. 2b.

### Two-photon verification

To validate the accuracy of our results, we compare them against those of two-photon microscopy, which are considered the ground truth. Figure 2b shows the results for a 160 μm thick sample of 4.8 μm green fluorescent beads. Miniscope3D accurately recovers all the beads in the volume after visually adjusting for tip/tilt misalignment in post-processing.

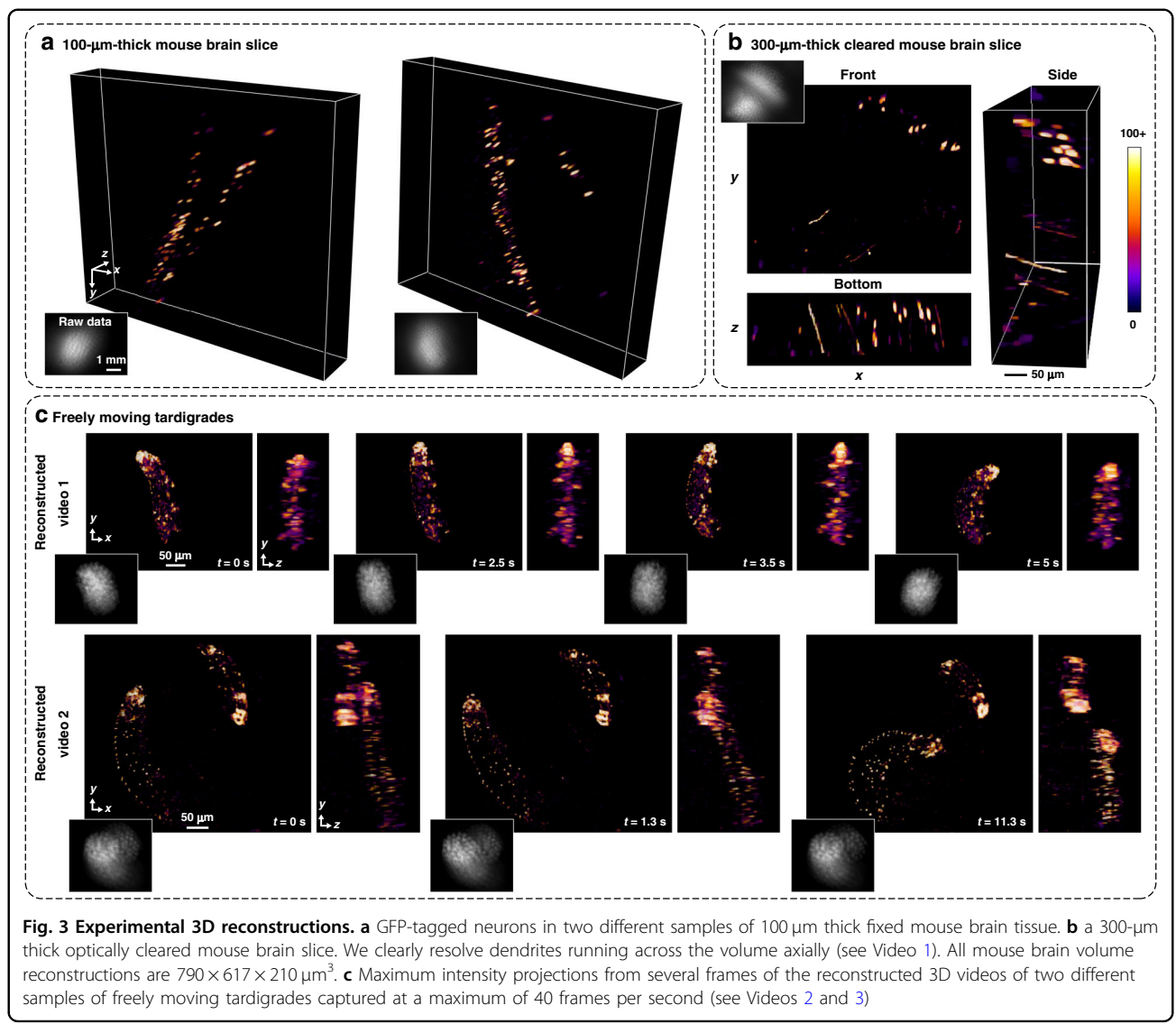

**Fig. 3 Experimental 3D reconstructions. a** GFP-tagged neurons in two different samples of 100 μm thick fixed mouse brain tissue. **b** a 300-μm thick optically cleared mouse brain slice. We clearly resolve dendrites running across the volume axially (see Video 1). All mouse brain volume reconstructions are $790 \times 617 \times 210\ \mu m^3$. **c** Maximum intensity projections from several frames of the reconstructed 3D videos of two different samples of freely moving tardigrades captured at a maximum of 40 frames per second (see Videos 2 and 3)

## Mouse brain tissue

Next, we show the feasibility of applying our design to neuro-biological samples by imaging post-fixed mouse brain slices, where the GFP is expressed in a sparse population of neurons throughout the sample. Figure 3a shows reconstructions from two 100 μm thick scattering samples from different parts of the hippocampus, and Fig. 3b shows the results from a 300 μm thick optically cleared section. In the 300 μm slice, dendrites can be seen running across the reconstruction axially (1 μm features), and individual cell bodies appear at distinct depths (see Video 1).

## Dynamic biological samples

Finally, we image dynamic samples of freely swimming SYBR-green stained tardigrades at a maximum of 40 frames per second. Figure 3c shows the maximum intensity projections of the reconstructed videos at different time points from two different recordings. The reconstructions show that Miniscope3D can track freely moving biological samples at high spatial and temporal resolution (see Videos 2, 3, 4, 5, 6).

## Discussion

Our device is designed with compressed 3D imaging and miniaturization in mind. For some 2D imaging applications where the losses in SNR (see Supplementary Fig. 6) and lateral resolution (2.76 μm vs 2 μm) are acceptable, our device may have advantages over the 2D Miniscope owing to its smaller size (17 mm vs. 23.5 mm tall) and weight (2.5 grams vs. 3 grams) or the ability to digitally refocus via 3D reconstruction. However, we expect that most applications of Miniscope3D will be for true 3D microscopy; thus, we mainly compare our specifications to the MiniLFM, which is considered the gold standard for single-shot miniature 3D fluorescence imaging.

Miniscope3D offers multiple improvements over the MiniLFM. First, using multifocal microlenses (as opposed to unifocal in an LFM) allows us to achieve better lateral resolution (2.76–3.9 μm) across a larger depth range (390 μm³). In contrast, the MiniLFM[19] demonstrated the best-case lateral resolution of 6 μm at a particular depth, and although the resolution at other depths was not reported, we predict that the unifocal microlens design will result in a lateral resolution that degrades significantly beyond 40 μm depth, based on previous analysis[17] and that in the "Multifocal Design" section below. We estimate that our Miniscope3D provides an ~10× increase in the usable measurement volume over the MiniLFM, with a 2.2× better peak lateral resolution. Taken together, our Miniscope3D reconstructs ~50× more usable voxels than the MiniLFM, significantly improving the utility of the device. This improved performance comes in a hardware package that is smaller than that of the MiniLFM (17 mm tall vs. 26 mm tall) and lighter weight (2.5 grams vs. 4.7 grams) because we replace the heavy doublet lens and the microlens array assembly with a thin phase mask. This arrangement will be particularly valuable in head-mounted experiments with freely moving animals.

Both our method and the MiniLFM make sparsity assumptions on the sample to solve the inverse problem to recover a 3D volume from a 2D image. We require the sample to be sparse in some domain, meaning that there is some representation of the sample that has many zeros in its coefficients[12,24]. Fluorescence imaging is a good candidate for these priors, as most biological samples are sparsely labeled. Because we optimize the microscope optics explicitly for single-shot 3D imaging, typical sparsity priors such as native sparsity, sparse 3D gradients (total variation (TV), as used in this paper), or sparse wavelets work well in our system. The MiniLFM is designed specifically for neural activity tracking and thus makes further structural and temporal sparsity assumptions, which improves the axial resolution from 30 μm (single-shot performance) to 15 μm (temporal video processing performance). In contrast, our Miniscope3D achieves a 15 μm single-shot axial resolution across a large depth range and could presumably improve upon that by incorporating temporal application-specific priors. In this paper, however, we aim to record highly dynamic samples (see supplementary videos) and thus impose only sample sparsity. We demonstrate the generality of our approach experimentally with samples that exhibit different levels of sparsity (Figs. 2 and 3), achieving resolution sufficient for single-neuron imaging. As sparsity decreases, the image quality and resolution degrade smoothly (see Supplementary Fig. 7), roughly following previous theoretical analyses[12,24,25].

Scattering is a limitation for all single-photon microscopes, including ours. For applications such as neural imaging and studying the 3D motion of freely swimming samples such as *Caenorhabditis elegans* or tardigrades, the small amount of scattering should not hinder the resolution. However, as the imaging depth within the scattering medium increases, we expect the resolution to degrade in a way similar to that of other single-photon microscopes. We show experimental reconstructions with and without scattering for the 100 μm thick scattering mouse brain tissue and the 300 μm thick cleared brain tissue. Both reconstructions achieve single-neuron resolution.

Another limitation of our model is that it assumes no partial occlusions. This is a common limitation of 3D recovery methods in fluorescence microscopy (e.g., double helix[26], light-field deconvolution microscopy[17], 3D localization microscopy) and generally works well in non-absorbing fluorescent samples. Modeling occlusions would be valuable in many practical situations but remains a challenging problem.

Accessibility was a key consideration in our Miniscope3D design. By building on the popular open-source Miniscope platform, our method can be easily adopted into existing experimental pipelines. Any of the 450 laboratories currently using the 2D Miniscope can upgrade to our 3D prototype with minimal effort. In addition, our method for 3D printing custom phase masks can enable others to fabricate their own mask designs tailored to particular applications. Because the experimental results are in good agreement with our theoretical design and analysis, we are confident that our design theory can serve as a useful framework for the future customization of single-shot 3D systems.

## Materials and methods
### System theory

Miniscope3D encodes volumetric information via a thin phase mask placed at the aperture stop of the gradient index (GRIN) objective lens (see Fig. 1). The goal of our design is to optimize the microscope optics for compressed sensing, enabling the capture of a large number of voxels from a small number of sensor pixels. To achieve this, the phase mask comprises an engineered pattern of multifocal microlenses, designed such that each fluorescent point source in the scene produces a unique high-frequency pattern of focal spots at the sensor plane, thus encoding its 3D position. The structure and spatial frequencies present in this pattern, termed the point spread function (PSF), determine our reconstruction resolution at that position; the theory for these limits is presented in the "Lateral Resolution" section below.

Figure 4 shows how our PSF changes with the lateral and axial position of a point source in the object space. As the point source moves laterally, the PSF translates (Fig. 4b). In an idealized microscope with the phase mask in Fourier space, the system would be shift invariant[21,22];

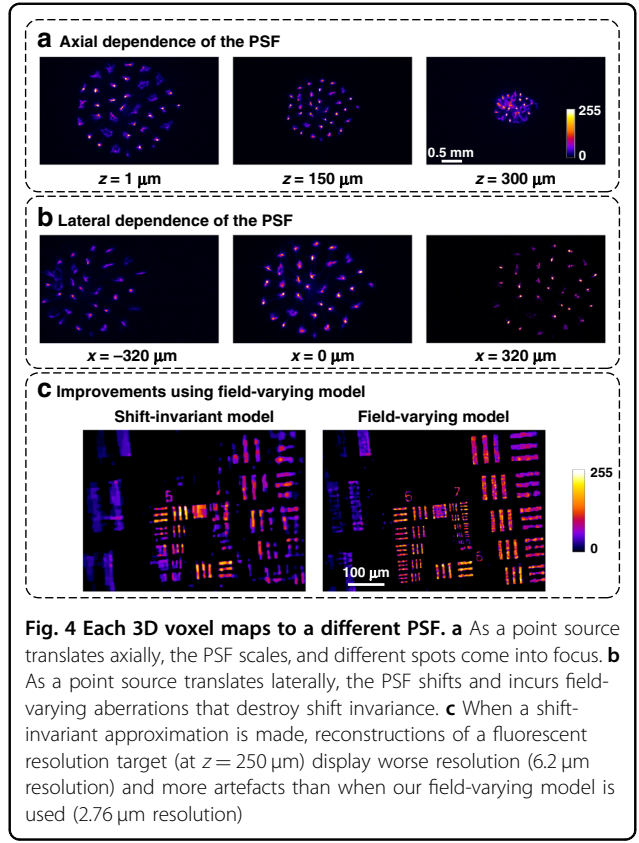

**Fig. 4 Each 3D voxel maps to a different PSF. a** As a point source translates axially, the PSF scales, and different spots come into focus. **b** As a point source translates laterally, the PSF shifts and incurs field-varying aberrations that destroy shift invariance. **c** When a shift-invariant approximation is made, reconstructions of a fluorescent resolution target (at $z = 250\,\mu m$) display worse resolution (6.2 μm resolution) and more artefacts than when our field-varying model is used (2.76 μm resolution)

however, because of the inherent aberrations in the GRIN lens, the pattern also slightly changes structure as it shifts. As the point source moves axially, the overall PSF changes size, and different spots come into focus (Fig. 4a) because we use a diversity of microlens focal lengths in our phase mask. As discussed in the section on "Multifocal Design," this ensures that the PSFs at a wide range of depths all contain sharp focal spots, unlike unifocal microlenses. To maximize the performance of our system, we optimize the spacing and focal lengths of the microlenses, as described in the "Phase Mask Optimization" section.

Our distributed, unique PSFs satisfy the multiplexing requirement of compressed sensing. Hence, we utilize sparsity-constrained inverse methods to recover the voxelized sparse 3D fluorescence emission, **v**, from a single 2D sensor measurement, **b**. To do this, we model **b** as a linear function of **v**, denoting the measurement process as **b** = $A$**v**. Here, $A$ is the measurement matrix, a linear operator that captures how each voxel maps to the sensor. Provided the sample is sparse in some domain, we reconstruct the volume by solving the sparsity-constrained inverse problem:

$$\hat{\mathbf{v}} = \arg\min_{\mathbf{v} \geq 0} \|A\mathbf{v} - \mathbf{b}\|_2^2 + \tau \|\Psi\mathbf{v}\|_1 \qquad (1)$$

with $\Psi$ being a sparsifying transform (e.g., 3D gradient, corresponding to TV regularization) and $\tau$ being a tuning parameter.

Equation (1) can be solved using a variety of iterative methods; we use the fast iterative shrinkage-thresholding algorithm (FISTA)[27]. This requires repeatedly applying $A$ and its adjoint. To make this computationally feasible for high megavoxel systems such as ours, we need an efficient representation for $A$. A shift-invariant forward model is extremely computationally efficient because $A$ becomes a convolution matrix[12,28,29]. It also requires only a single PSF calibration image, from which the PSFs at all other positions can be inferred. Unfortunately, miniature integrated systems such as ours are not shift invariant owing to the off-axis aberrations inherent to compact objectives. To account for this, in the following sections, we develop a field-varying forward model and a practical calibration scheme that account for aberrations with minimal added computational cost.

**Field-varying forward model**

Because aberrations in the GRIN lens of the Miniscope render the shift-invariant model invalid, we need to both measure and model how the PSF changes across the FoV. Explicitly measuring the PSF at each position within the volume is infeasible, both in terms of the amount of calibration data and the computational burden of reconstruction. It is also unnecessary as the PSF structure changes slowly across the FoV. Instead, our calibration scheme samples the PSF sparsely across the field and uses a weighted convolution model to estimate the PSF at other positions[30]. We capture 64 PSF measurements at each depth and then use them to predict the full set of over 300,000 PSFs. Our forward model thus requires only computing a limited number of convolutions (typically 10–20) and achieves a 2.2× better resolution and better quality than those of the shift-invariant model (see Fig. 4c).

Our field-varying forward model approximates $A$ using a weighted sum of shift-invariant (convolution) kernels. We treat the volumetric intensity as a 3D grid of voxels, denoted as $\mathbf{v}[x, y, z]$. A voxel at location $[x, y, z]$ produces a PSF on the sensor, $\mathbf{h}[u, v; x, y, z]$, where $[u, v]$ indexes sensor rows and columns. For ease of notation, we assume that the system has magnification $M = 1$ and apply appropriate scaling to the solution after 3D image recovery. We also assume that **v** has finite axial and lateral support. By treating the voxels as mutually incoherent, the measurement becomes a linear combination of PSFs:

$$\mathbf{b}[u, v] = \sum_z \sum_{x,y} \mathbf{v}[x, y; z]\mathbf{h}[u, v; x, y, z]$$
$$= A\mathbf{v} \qquad (2)$$

where the bounds of the sums implicitly contain the sample. To capture the field-varying behavior, we seek to model the PSF from each voxel as a weighted sum of $K$ shift-invariant kernels[30]. The kernels, $\mathbf{g}_r[u, v; z]$, and weights, $\mathbf{w}_r[x, y, z]$, which are described below, should be chosen to represent all PSFs accurately with the smallest possible $K$. Mathematically, the forward model can be written as:

$$\mathbf{h}[u, v; x, y, z] = \Lambda[u, v] \sum_{r=1}^{K} \mathbf{w}_r[x, y, z] \mathbf{g}_r[u - x, v - y; z] \tag{3}$$

where $\Lambda[u, v]$ is an indicator function that selects only the values that fall within the sensor pixel grid. In other words, the PSF from position $[x, y, z]$ is modeled by shifting the kernels, $\{\mathbf{g}_r[u, v; z]\}$ $r = 1 \dots K$, associated with depth $z$, to be centered at the PSF location on the sensor, $[u, v] = [x, y]$. Then, each kernel is assigned a field-dependent weight, $\mathbf{w}_r[x, y, z]$, and the weighted kernels are summed over $r$. Note that this motivates the placement of the phase mask in the aperture stop. Because all field points fully illuminate the mask, the system becomes nearly shift-invariant, which keeps the necessary number of kernels low.

To find the kernels and weights that best represent all of the PSFs, we first consider each PSF in a coordinate space relative to the chief ray. We do this by centering each measured PSF on-axis. We do this by centering each measured PSF on-axis:

$$\mathbf{h}[u + x, v + y; x, y, z] = \sum_{r=1}^{K} \mathbf{w}_r[x, y, z] \mathbf{g}_r[u, v] \tag{4}$$

where $[x, y]$ is the chief ray spatial coordinate at the sensor. We assume that the calibration procedure will capture $N$ PSFs across the field, $\{\mathbf{h}[u, v; x_i, y_i, z]\}$ $i = 1 \dots N$, for each depth $z$. We estimate the chief ray coordinate $[x, y]$ of off-axis PSFs by cross-correlating each with the on-axis PSF. The off-axis measurements are then shifted on-axis, vectorized, and combined into a registered PSF matrix, denoted as $H$. For smoothly varying systems, $H$ is low rank and can be well approximated by solving

$$\hat{G}, \hat{W} = \underset{G, W}{\arg\min} \|GW - H\|_2^2 \tag{5}$$

where $G \in \mathbb{R}^{M_p \times K}$ and $W \in \mathbb{R}^{K \times N}$ for a sensor with $M_p$ pixels. The optimal rank-$K$ solution can be found by computing the $K$ largest values of the singular value decomposition of $H$. The $r$th column of the left singular vector matrix, $\hat{G}$, contains the kernel $\mathbf{g}_r[x, y; z]$ in vectorized form. Similarly, combining the singular values with the right singular vector matrix produces $\hat{W}$, of which the $r$th row contains the optimal weights $\mathbf{w}_r[x_i, y_i, z]$ for voxel $[x_i, y_i, z]$. Empirically, we find that the weights vary smoothly across the field; hence, we use natural neighbor interpolation to estimate the weights between sampled points. After empirically testing the number of sample points per depth ($N$), we find 64 to be sufficient for our system.

The computational efficiency of this model can be analyzed by substituting Eq. (3) into Eq. (2), yielding:

$$\mathbf{b}[u, v] = \sum_z \sum_{x, y} \mathbf{v}[x, y, z] \Lambda[u, v] \sum_{r=1}^{K} \mathbf{w}_r[x, y, z] \mathbf{g}_r[u - x, v - y; z]$$

$$= \Lambda[u, v] \sum_z \sum_{r=1}^{K} \left\{ (\mathbf{v}[x, y, z] \mathbf{w}_r[x, y, z]) \overset{[x, y]}{*} \mathbf{g}_r[x, y; z] \right\}[u, v] \tag{6}$$

where $\overset{[x, y]}{*}$ denotes discrete linear convolution over the lateral variables. In practice, each convolution can be implemented using a combination of padding and FFT convolution, whereas $\Lambda[u, v]$ represents a crop[12]. Note that the summation over $z$ assumes that no voxel is partially occluded. Because this model comprises $K$ point-wise multiplications and $K$ 2D convolutions per depth, it is approximately $K$–times slower than a shift-invariant model. Hence, minimizing $K$ via the choice of weights and kernels or by reducing aberrations in the hardware improves the computational efficiency.

## Calibration

Experimentally, our calibration procedure captures PSF images of a 2.5 μm green fluorescent bead at 64 equally spaced points across the FoV for each depth. Empirically, we find that the singular values decay quickly and that a model with rank between $K = 10$ and $K = 20$ is sufficient for our system. Note that we can trade-off the speed and accuracy of our model by varying $K$, but the decomposition needs to be performed only once. This method allows characterization of an extremely large matrix by capturing only a relatively small number of images. For example, our typical calibration requires 80 depths. Densely sampling every PSF using a 0.3 megapixel sensor would require 24 million calibration images (300,000 per depth) and terabytes of storage. In contrast, our method enables calibrating this entire volume using only 80 depths × 64 images/depth = 5120 images, which takes 2 h to capture using automated stages and requires a few gigabytes to store.

## Reconstruction algorithm

In solving Eq. (1), we use the sparsifying transform $\Psi = [\nabla_x \nabla_y \nabla_z]^T$, which corresponds to 3D anisotropic TV regularization, promoting sparse 3D gradients in the reconstruction. The regularization parameter, $\tau$, controls the balance between the data fidelity and the sparse 3D gradients prior. In practice, we hand-tune $\tau$ on a range of test data and then leave it fixed for subsequent captures (see Supplementary Fig. 5). We solve Eq. (1) using the FISTA[27] with the fast, subiteration-free parallel proximal method[31]. Computationally, our method has similarities to light-field deconvolution[17], but because our PSF is not periodic and our focal lengths are not all the same, we are able to remove the need for aperture matching and achieve higher resolution across a larger volume. To solve Eq. (1), we compute the linear forward and adjoint matrix-vector multiples using an FFT convolution. A typical reconstruction takes 1–3 k iterations and runs in 8–24 min on a GPU RTX 2080-Ti using MATLAB.

## Phase mask design

In this section, we present the theory for designing and optimizing a phase mask that achieves a target resolution uniformly across a specified 3D volume. We assume that the phase mask will be placed in the aperture stop of the objective with the sensor at a fixed distance, as this architecture reduces the size and weight of our device, makes the system close to shift-invariant and enables multiplexing, which is necessary for compressed sensing. We aim for all PSFs produced by the mask to have high spatial-frequency content and be mutually incoherent (i.e., all as dissimilar as possible). Toward this goal, we propose a multifocal array of nonuniformly spaced microlenses as our phase mask.

We choose to use a phase mask made of microlenses because it provides good light throughput while balancing the trade-offs between the SNR and compressive sensing capabilities. Our previous work employed off-the-shelf diffusers with a pseudorandom Gaussian surface profile[12]. These generate a caustic PSF that has poor SNR owing to the spreading of the light by the concave bumps of the diffuser surface. In contrast, microlenses concentrate light into a small number of sharp spots, giving better performance in low-light applications such as fluorescence microscopy (see Supplementary Sec. 1). By parameterizing our design as a set of microlenses, we can also derive simple design rules from first principles (sections "Lateral Resolution & Multifocal Design") and then use them to formulate an optimization problem that locally optimizes the placement and aberrations of each microlens.

We space our microlenses nonuniformly to ensure that the PSFs from all field points are dissimilar. Regularly spaced arrays will produce highly similar PSFs when shifted by one microlens period, causing certain spatial frequencies to be poorly measured. Previous work avoided this ambiguity by introducing a field stop[21–23] that prevents the PSFs from overlapping, but this significantly restricts the FoV. Our design yields a larger FoV by using nonuniform spacing and computationally disambiguating the overlapping PSFs. In Fig. 5, we compare PSFs and reconstructions from regularly spaced and nonuniform phase mask designs. Looking at Fig. 5c, the PSF of the regular array causes unwanted peaks at low frequencies in its radially averaged inverse power spectral density (IPSD), a metric related to deconvolution performance[32] (lower is better). This manifests as artefacts in the simulated reconstruction, which are significantly reduced in reconstructions from both of the nonuniform designs.

Using multiple microlens focal lengths extends the depth range across which we obtain good resolution, as described in the section on "Multifocal Design". Multifocal designs have sharp focal spots across a wider desired depth range than can be achieved with unifocal designs, trading SNR in-focus for better performance off-focus. Figure 5c, d compares the PSFs and reconstruction quality of our approach versus those of unifocal designs in-focus and 200 μm away from the native focus of the unifocal arrays. The blurry features in the out-of-focus PSFs for both unifocal designs cause poor performance, as shown in the reconstructions and high inverse power spectra. To capture the performance across multiple depths, Fig. 5b shows the integrated IPSD (up to the cutoff frequency) of each design versus depth. As expected, our multifocal design is slightly worse than a unifocal design in focus but achieves far better (lower) values across the full depth range.

In the compact system architecture we propose, it is clear that our nonuniform multifocal microlenses are a good choice of phase mask. This fact motivates the next sections, which provide guidance on optimizing the nonuniform spacing as well as the focal lengths and aberrations of the microlenses for achieving a target resolution and depth range. For our prototype, we aim for a 3.5 μm lateral resolution and show that this can be achieved over a depth range up to 360 μm, which agrees with our experimental characterization.

## Lateral resolution

The lateral resolution is primarily determined by the diffraction-limited aperture size of the microlenses, which also determines the number of microlenses that fit across the objective's full aperture and thus the depth range we can target. We design a lateral resolution that does not require the full pupil so that we can fit multiple microlenses in the aperture for better depth coding. The example in Fig. 5 targets a 3.5 μm resolution (cutoff frequency of 0.35 cycles/μm) using 36 microlenses with an average NA = 0.09. Because each design has the same number of microlenses, each has a similar resolution limit.

To quantify, we perform a diffraction analysis to find the clear aperture a single microlens needs to support a $\delta x$ lateral resolution at the sample. Note that this assumes that we will recover resolution no better than the bandlimit of the measurement, neglecting any resolution gained from the non-linear solver. We start by calculating the magnification of our system:

$$M \approx \frac{-t}{f_G} \tag{7}$$

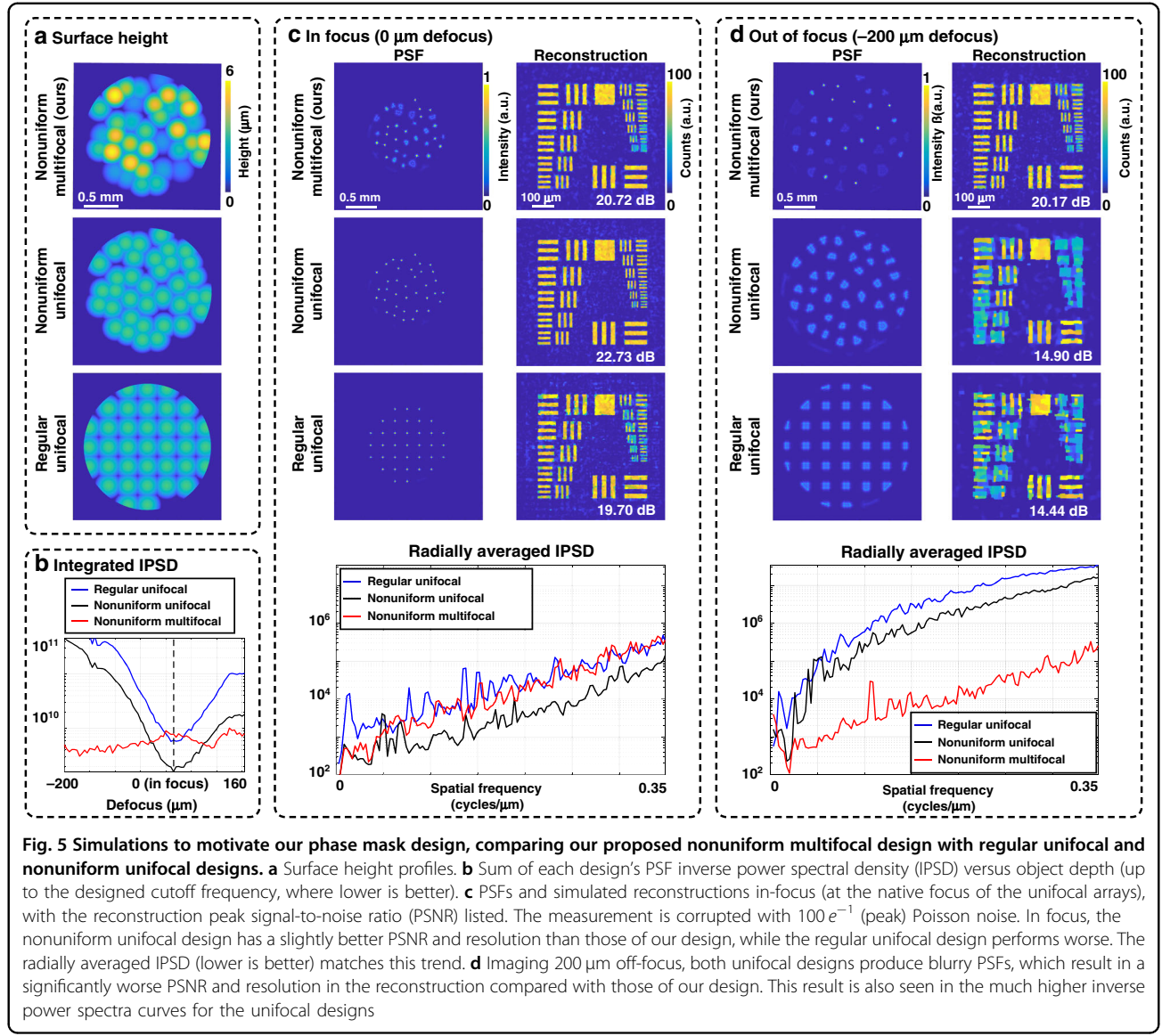

**Fig. 5 Simulations to motivate our phase mask design, comparing our proposed nonuniform multifocal design with regular unifocal and nonuniform unifocal designs. a** Surface height profiles. **b** Sum of each design's PSF inverse power spectral density (IPSD) versus object depth (up to the designed cutoff frequency, where lower is better). **c** PSFs and simulated reconstructions in-focus (at the native focus of the unifocal arrays), with the reconstruction peak signal-to-noise ratio (PSNR) listed. The measurement is corrupted with $100\,e^{-1}$ (peak) Poisson noise. In focus, the nonuniform unifocal design has a slightly better PSNR and resolution than those of our design, while the regular unifocal design performs worse. The radially averaged IPSD (lower is better) matches this trend. **d** Imaging 200 µm off-focus, both unifocal designs produce blurry PSFs, which result in a significantly worse PSNR and resolution in the reconstruction compared with those of our design. This result is also seen in the much higher inverse power spectra curves for the unifocal designs

where $f_G$ is the GRIN focal length and $t$ is the mask-to-sensor distance (derivation in Supplementary Sec. 3). Note that $M$ is approximately independent of the microlens focal length. For our system, $f_G = 1.67$ mm and $t = 8.7$ mm; thus, $M \approx -5.2$. Using Eq. (7) and the Rayleigh criterion, the microlens clear aperture, $\Delta_{ML}$, needed for a target object resolution $\delta x$ at wavelength $\lambda$ is:

$$\Delta_{ML} = \frac{1.22\lambda t}{|M|\delta x} \approx \frac{1.22\lambda f_G}{\delta x} \quad (8)$$

This expression is also independent of the microlens focal length because we have assumed that the microlens is focused. Equation (8) allows us to select the appropriate average microlens spacing for a desired resolution. Our system is designed for a 3.5-µm lateral resolution (though experimentally, we achieve 2.76 µm owing to the non-linear solver), which gives an average microlens diameter of 300 µm. Given that the GRIN clear aperture has a diameter of 1.8 mm, this results in 36 microlenses that can fit in the phase mask. Note that since the GRIN is aberration limited, the 2D Miniscope does not achieve the diffraction-limited resolution predicted by its full aperture size. Hence, our experimentally measured resolution is not much worse than that of the 2D Miniscope (lateral resolution of 2 µm), despite dividing the GRIN pupil into 36 regions to add depth-sensing capabilities.

**Multifocal design for extended depth range**

Focal length diversity in the microlens array results in an extended depth range, a key advantage of our architecture over the conventional LFM. To maintain a uniform lateral resolution across all depths in the volume of interest, the PSF should have sharp, high-frequency focal spots for each axial position. This requires at least one microlens to be in focus for each object axial plane, with planes spaced by the microlens depth-of-field (DoF). The DoF of a single microlens, $d_{ML}$, is inversely proportional to the microlens clear aperture, $\Delta_{ML}$, giving $d_{ML} = \pm 20\,\mu m$ for our system (see Supplementary Sec. 4 for details).

Our design aims to have a minimum of four microlenses in focus within each DoF. Given that our lateral resolution criterion allows 36 microlenses, we should have nine different focal lengths and a depth range of 360 μm, ~10× what a single focal length achieves. Note that there is a trade-off between the imaging depth range and lateral resolution. We can increase the depth range by including more microlenses; however, doing so decreases their clear aperture (Eq. (8)) and thus the lateral resolution. Conversely, for imaging thin samples where only a narrow range of focal lengths is required, better lateral resolution is possible.

To determine the focal length distribution, we find the focal length needed to focus at the beginning of the depth range ($f_{min} = 7\,mm$) and at the end of the depth range ($f_{max} = 25\,mm$). Then, we dioptrically space the focal lengths across the target range because this leads to microlenses that come into focus at linearly spaced depth planes in the sample space.

**Phase mask parameterization**

The previous sections outlined the first-order design principles, considering only a single microlens. In the next section, we optimize the ensemble of microlenses (their positions and added aberrations) with metrics based on compressed sensing theory. Here, we first build our representation of the microlens phase mask by parameterizing the $i$th microlens by its lateral vertex location, $(\rho_{xc}^i, \rho_{yc}^i) := \boldsymbol{\rho}_c^i$, and radius of curvature, $R_i$. The spherical sag of the microlens is:

$$s_i = d_i + R_i \sqrt{1 - \left(\frac{\boldsymbol{\rho} - \boldsymbol{\rho}_c^i}{R_i}\right)^2} \qquad (9)$$

where $d_i$ is an offset constant added to each microlens to control its clear aperture. We parameterize aspheric terms in the microlenses by using Zernike polynomials. The $j$th Zernike coefficient for microlens $i$ is denoted by $\alpha_{ij}$; thus, the total aspheric component at that microlens is $\sum_j \alpha_{ij} Z_j(\boldsymbol{\rho} - \boldsymbol{\rho}_c^i)$, with $Z_j$ being the $j^{th}$ Zernike polynomial. As long as the microlenses are all convex ($R_i > 0$), a phase mask with a full fill-factor can be constructed by taking the point-wise maximum thickness (see Fig. 6). The phase

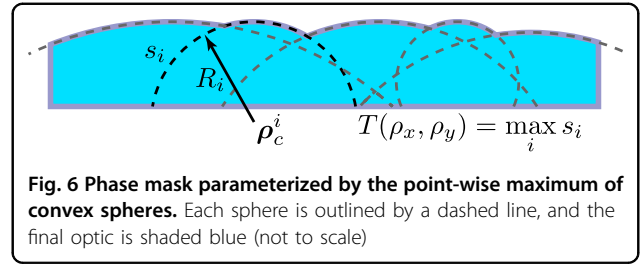

**Fig. 6 Phase mask parameterized by the point-wise maximum of convex spheres.** Each sphere is outlined by a dashed line, and the final optic is shaded blue (not to scale)

mask surface is thus:

$$T(\rho_x, \rho_y; \boldsymbol{\theta}) = \max_i \left[ s_i + \sum_j \alpha_{ij} Z_j(\boldsymbol{\rho} - \boldsymbol{\rho}_c^i) \right] \qquad (10)$$

where $\boldsymbol{\theta}$ denotes the collection of parameters that define the phase mask: vertex locations $\{\boldsymbol{\rho}_c^i\}$, radii $\{R_i\}$, offsets $\{d_i\}$, and Zernike coefficients $\{\alpha_{ij}\}$. The resulting surface is guaranteed to be continuous and will have a well-defined local focal length given by $f_i = \frac{n-1}{R_i}$ within the region belonging to the $i$th microlens, provided the power Zernike $j = 4$ is excluded. In practice, we optimize the Zernike coefficients for tilt ($j = 1, 2$) and astigmatism ($j = 3, 5$).

With the microlens array defined, the on-axis PSF at a given sample depth $z$ can be modeled by Fresnel propagation of the pupil wavefront from a point source at depth $z$, denoted by $W(\rho_x, \rho_y; z)$, multiplied by the phase of the designed mask, $\phi(\rho_x, \rho_y; \boldsymbol{\theta}) = \frac{2\pi(n-1)}{\lambda} T(\rho_x, \rho_y; \boldsymbol{\theta})$:

$$\mathbf{h}(u, v; z, \boldsymbol{\theta}) = \left| F_t \left\{ P(\rho_x, \rho_y) \exp \left[ i\phi(\rho_x, \rho_y; \boldsymbol{\theta}) \right] W(\rho_x, \rho_y; z) \right\} \right|^2 \qquad (11)$$

where $P(\rho_x, \rho_y)$ is the GRIN pupil amplitude, $n$ is the microlens substrate index of refraction, and $F_t$ denotes Fresnel propagation to the sensor a distance $t$ away. Importantly, the on-axis PSFs are differentiable with respect to the microlens parameters, $\boldsymbol{\theta}$, enabling us to optimize the design using gradient methods, as discussed in the next section.

**Phase mask optimization using matrix coherence**

Given the first-principles guidance in the above sections, we set the number of microlenses, their characteristic aperture size and their focal length distribution; next, we aim to optimize the microlens positions and aberrations to maximize the performance. To make the optimization computationally feasible, we ignore the field-varying changes in the PSF and assume that the system is shift invariant for the purposes of design.

To optimize the microlens parameters, $\boldsymbol{\theta}$, in terms of the on-axis PSFs at each depth, we set up a loss function to be optimized that consists of two terms. The first term, a cross-coherence loss, promotes good axial resolution by ensuring that the PSFs at different depths are as dissimilar as possible. Cross-coherence between any two depths is defined as $\| \mathbf{h}(u, v; z_n) \star \mathbf{h}(u, v; z_m) \|_\infty := \max[\mathbf{h}(u, v; z_n) \star \mathbf{h}(u, v; z_m)]$, where $\star$ represents the 2D correlation and max [·] is the element-wise maximum. Intuitively, we want the cross-coherence to be small, as it represents the worst-case ambiguity that would arise by placing two point sources adversarially at depths spaced according to the separation of their PSF cross-correlation peaks. By computing this quantity for all pairs of $z$-depths, we can produce a differentiable figure-of-merit that optimizes the matrix coherence[24] between depths. In practice, rather than optimizing the cross-coherence, we smoothly approximate the max[33] using $\|x\|_\infty \approx \sigma \ln \sum \exp(x^2/\sigma)$. Here, $\sigma > 0$ is a tuning parameter that trades off the accuracy of the approximation with the smoothness. For our purposes, this has the advantage of penalizing all large cross-correlation values, not just the single largest. We denote this $\| \cdot \|_{\overline{\infty}}$.

The total cross-coherence loss is then:

$$q(\boldsymbol{\theta}) = \sum_n \sum_{m>n} \|\mathbf{h}(u, v; \boldsymbol{\theta}, z_n) \star \mathbf{h}(u, v; \boldsymbol{\theta}, z_m)\|_{\overline{\infty}} \tag{12}$$

The second term in the optimization ensures that the lateral resolution is maintained. To do so, we optimize the autocorrelation of the PSF at each depth using the frequency domain least-squares method. The analysis in the "Lateral Resolution" section above applies only to a single microlens; building a phase mask of multiple lenses generally degrades resolution by introducing dips in the spectrum that reduce contrast at certain spatial frequencies. Hence, we treat the single-lens case as an upper limit that defines the bandlimit of the multi-lens PSF. To reduce the spectral ripple, we penalize the $\ell_2$ distance between the MTFs of the PSF and a diffraction-limited single microlens, $|H|$. We include a weighting term, denoted as $D$, that ignores spatial frequencies beyond the bandlimit, as well as low spatial frequencies that are less critical and difficult to optimize owing to out-of-focus microlenses. The autocorrelation design term is then:

$$p(\boldsymbol{\theta}) = \sum_n \left\| D \left[ \mathbb{F} \{ \mathbf{h}(u, v; \boldsymbol{\theta}, z_n) \star \mathbf{h}(u, v; \boldsymbol{\theta}, z_n) \} - |H|^2 \right] \right\|_2^2 \tag{13}$$

where $\mathbb{F}\{\cdot\}$ is the 2D discrete Fourier transform.

The total loss is the weighted sum of the two terms:

$$f(\boldsymbol{\theta}) = p(\boldsymbol{\theta}) + \tau_0 q(\boldsymbol{\theta}) \tag{14}$$

where $\tau_0$ is a tuning parameter used to control their relative importance. To initialize, we randomly generate 5000 heuristically designed candidate phase masks, each with 36 microlenses spaced according to Poisson disc sampling across the GRIN aperture stop. The focal lengths are distributed dioptrically between the minimum and maximum values computed from the "Multifocal Design" section. The best candidate from these 5000 is then optimized using gradient descent applied to $f\{\boldsymbol{\theta}\}$. This process is implemented in TensorFlow Eager to enable GPU-accelerated automatic differentiation.

The results of our optimized design are shown in Fig. 7, where we compare our optimized mask to the random multifocal design that scored worst during initialization

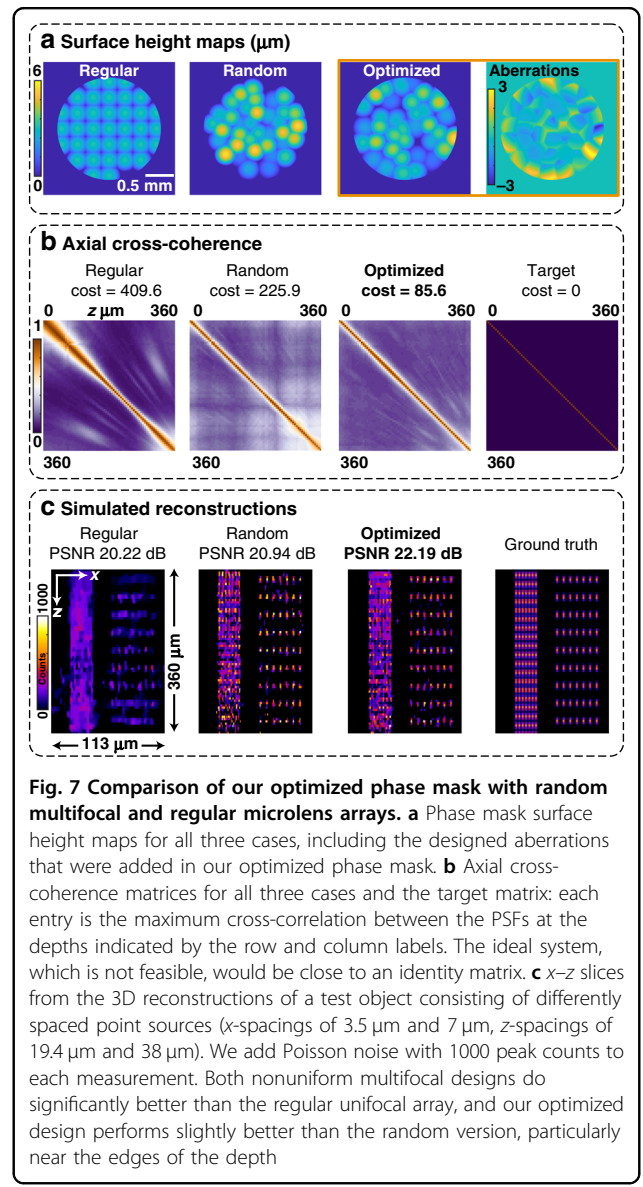

**Fig. 7 Comparison of our optimized phase mask with random multifocal and regular microlens arrays. a** Phase mask surface height maps for all three cases, including the designed aberrations that were added in our optimized phase mask. **b** Axial cross-coherence matrices for all three cases and the target matrix: each entry is the maximum cross-correlation between the PSFs at the depths indicated by the row and column labels. The ideal system, which is not feasible, would be close to an identity matrix. **c** $x$–$z$ slices from the 3D reconstructions of a test object consisting of differently spaced point sources ($x$-spacings of 3.5 μm and 7 μm, $z$-spacings of 19.4 μm and 38 μm). We add Poisson noise with 1000 peak counts to each measurement. Both nonuniform multifocal designs do significantly better than the regular unifocal array, and our optimized design performs slightly better than the random version, particularly near the edges of the depth

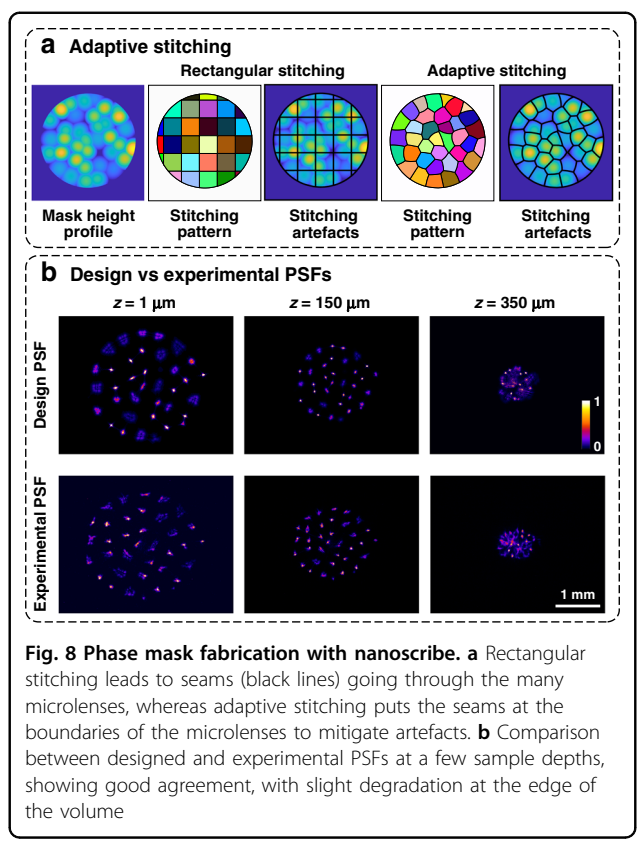

**Fig. 8 Phase mask fabrication with nanoscribe. a** Rectangular stitching leads to seams (black lines) going through the many microlenses, whereas adaptive stitching puts the seams at the boundaries of the microlenses to mitigate artefacts. **b** Comparison between designed and experimental PSFs at a few sample depths, showing good agreement, with slight degradation at the edge of the volume

and a regular unifocal array. The optimized design has the best axial cross-coherence (Fig. 7b), with the random array having worse off-diagonal terms. Hence, in the 3D reconstructions (Fig. 7c), the optimized design performs slightly better than the random design. The regular microlenses produce large off-diagonal peaks in the cross-coherence, which manifests as poor 3D reconstruction performance off-focus.

## Phase mask fabrication

As our phase mask designs can be tailored to specific applications with different resolution requirements and volumes of interest, the ability to rapidly generate phase mask prototypes is very useful. Recently, the Nanoscribe two-photon polymerization 3D printer has been shown to print free-form microscale optics on-demand[34]. However, in its current implementation, Nanoscribe uses planar galvanometric scanning to polymerize the resist, resulting in a limited FoV (diameter of ~350 μm with the ×25 Nanoscribe objective). If larger objects need to be printed, several blocks need to be stitched together by moving the substrate with a mechanical stage. Stitching artefacts from this process can seriously impact the produced object[35], usually by causing rectangular or hexagonal blocking artefacts. As seen in Fig. 8a, rectangular seams going through the center of the microlenses can be very detrimental to our design.

One solution to this is an adaptive stitching algorithm that has been demonstrated for slender objects and a non-overlapping microlens array[35]. Here, we propose a new height-based segmentation algorithm capable of placing the stitching seams in the overlapping region between the overlapping microlenses (Fig. 8a). This is based on the local height functions for each microlens, described in the "Phase Mask Parameterization" section. Each of these functions has a region where they result in the largest values, and this region is precisely the printing block that will be printed from that microlens center location (see Supplementary Section 10). Once the adaptive stitching mask is obtained, the writing instructions per block can be generated using TipSlicer[36,37]. Figure 8b compares the designed and experimental PSFs at three depth planes, showing a good match with some degradation at the end of the volume.

## Device assembly

Our prototype Miniscope3D system consists of a custom phase mask, a CMOS sensor (Ximea MU9PM-MH), a fluorescent filter set (Chroma ET525/50m, T495lpxr, ET470/40x), a GRIN lens (Edmund Optics 64–520), and a half-ball lens (Edmund 47–269), with a 3D-printed optomechanical housing. The 55 μm thick phase mask is glued to the back surface of the GRIN lens using optical epoxy. Note that our experimental PSF calibration accounts for slight misalignment in the phase mask. The final device is 17 mm tall and weighs 2.5 grams.

### Acknowledgements
This work was supported in part by the Defense Advanced Research Projects Agency (DARPA), contract no. N66001-17-C-4015, Gordon and Betty Moore Foundation Data-Driven Discovery Initiative (grant GBMF4562), National Institutes of Health (NIH) grant 1R21EY027597-01, the National Science Foundation (grant no. 1617794), and an Alfred P. Sloan Foundation fellowship. Kyrollos Yanny acknowledges funding from the National Science Foundation Graduate Research Fellowship Program (NSF GRFP). We thank the UCLA Miniscope open-source team for their technical advice and support. We also thank Hannah Gemrich and Dr. Saul Kato for providing the tardigrade samples.

### Author details
[1]UCB/UCSF Joint Graduate Program in Bioengineering, University of California, Berkeley, CA 94720, USA. [2]Department of Electrical Engineering & Computer Sciences, University of California, Berkeley, CA 94720, USA. [3]TIPs Department, Université libre de Bruxelles (ULB), 1050 Brussels, Belgium

### Author contributions
K.Y. and N.A. designed the microscope; K.Y., N.A., and W.L. planned and conducted the experiments; K.Y., N.A., and W.L. analyzed the data; L.W. and R.N. supervised the project; K.Y., N.A., W.L., and L.W. wrote the manuscript; and all coauthors contributed to the manuscript.

### Conflict of interest
The authors declare that they have no conflict of interest.